\documentclass[]{spie}  

\usepackage{amsmath,amsfonts,amssymb}
\usepackage{graphicx}
\usepackage[colorlinks=true, allcolors=blue]{hyperref}
\usepackage{multirow}
\usepackage{orcidlink}

\title{Correlating Visual Characteristics and Cryogenic Performance of Superconducting Detectors}

\author[a,b]{K.~R.~Ferguson\,\orcidlink{0000-0002-4928-8813}}
\author[b,c,d]{A.~N.~Bender\,\orcidlink{0000-0001-5868-0748}}
\author[e]{N.~Whitehorn\,\orcidlink{0000-0002-3157-0407}}
\author[b]{T.~W.~Cecil\,\orcidlink{0000-0002-7019-5056}}
\affil[a]{Department of Physics and Astronomy, University of California, Los Angeles, CA, 90095, USA}
\affil[b]{High-Energy Physics Division, Argonne National Laboratory, 9700 South Cass Avenue., Lemont, IL, 60439, USA}
\affil[c]{Department of Astronomy and Astrophysics, University of Chicago, 5640 South Ellis Avenue, Chicago, IL, 60637, USA}
\affil[d]{Kavli Institute for Cosmological Physics, University of Chicago, 5640 South Ellis Avenue, Chicago, IL, 60637, USA}
\affil[e]{Department of Physics and Astronomy, Michigan State University, East Lansing, MI 48824, USA}

\authorinfo{Corresponding author: K.~R.~Ferguson\\E-mail: \href{mailto:kferguson@physics.ucla.edu}{kferguson@physics.ucla.edu}}

\begin{document}

\newcommand{\fixme}[1]{\textcolor{red}{FIXME: #1}}

\maketitle

\begin{abstract}
Cryogenic characterization of transition-edge sensor (TES) bolometers is a time- and labor-intensive process. As new experiments deploy larger and larger arrays of TES bolometers, the testing process will become more of a bottleneck. Thus it is desirable to develop a method for evaluating detector performance at room temperature. One possibility is using machine learning to correlate detectors' visual appearance with their cryogenic properties. Here, we use three engineering-grade TES bolometer wafers from the production cycle for SPT-3G, the current receiver on the South Pole Telescope, to train and test such an algorithm. High-resolution images of these TES bolometers were captured and relevant features were calculated from the images. Cryogenic performance metrics, including a detector's ability to tune and superconducting parameters such as normal resistance, critical temperature, and transition width, were also measured. A random forest algorithm was trained to predict these performance metrics from the visual features. Analysis of the images proved highly successful. While the ability of the random forest algorithm to predict cryogenic features was limited with the chosen set of input image features, it is possible that an increase in data volume or the addition of more image features will solve this problem.
\end{abstract}


\section{Introduction}
\label{sec:intro}
Transition-edge sensor (TES) bolometers are widely used in state-of-the-art particle physics experiments as well as millimeter-wave and x-ray astronomical observations. When a bolometer absorbs energy, its temperature changes, prompting a change in its resistance. This change in resistance is read out to measure the signal on the sky. TES bolometers have superconducting thermistors that are voltage biased to operate precisely in their superconducting transition; thus, a small change in temperature yields a large change in resistance, enabling the detectors to be highly sensitive even to weak signals \cite{richards94, irwin05}.

The experiments enabled by TES bolometers touch on some of the largest questions in cosmology and particle physics today. For example, experiments such as SPT-3G (the current receiver on the South Pole Telescope [SPT]), BICEP Array, Advanced ACTPol (the current receiver on the Atacama Cosmology Telescope [ACT]), POLARBEAR/Simons Array, and Simons Observatory use TES bolometers to precisely map the temperature and polarization anisotropies of the cosmic microwave background (CMB), the relic radiation leftover from the early universe \cite{sobrin22, moncelsi20, crowley18, westbrook18, simons19}. SuperCDMS-Soudan uses TES bolometers to search for scattering of weakly-interacting massive particles (WIMPs) off regular matter, and the SCUBA-2 experiment maps cold gas and dust in the universe using TES bolometers \cite{supercdms18, holland13, mairs21}. Together, these experiments seek to determine the nature of dark matter and dark energy, to measure the precise value of the effective number of relativistic species, and to probe the universe's inflationary period \cite{abazajian15b, weinberg13, lesgourgues06, guth80, lyth96, abazajian15a, bucher15}.

For CMB experiments, the dominant source of noise for TES bolometers is Poisson noise from photon counting statistics. This noise decreases as more photons are collected; while more photons could be collected by increasing the size of the detectors, this would degrade the angular resolution of the telescopes. Thus in order to build more sensitive experiments, larger arrays of TES bolometers must be deployed. For example, current-stage CMB experiments like SPT-3G and Advanced ACTPol deploy $\mathcal{O}\left( 10{,}000 \right)$ detectors, while the upcoming CMB-S4 experiment is expected to deploy $\mathcal{O}\left( 500{,}000 \right)$ to reach its science goals \cite{abazajian16, abitbol17, abazajian19}. Characterization and verification of bolometer performance is a time- and labor-intensive process. Each wafer of $\mathcal{O}\left( 1{,}000 \right)$ bolometers must be packaged with readout electronics into modules and installed into a cryogenic chamber. The chamber must cool the detectors to sub-Kelvin temperatures after which performance data can be captured. The electrical and thermal properties of a TES bolometer include (but are not limited to) resistance $R$, critical temperature $T_\textrm{c}$, thermal conductance $G$, and saturation power $P_\textrm{sat}$. Optical properties such as the bandpass, efficiency, and polarization sensitivity can also be measured during this stage. Typically, this testing process takes a minimum of two weeks, without much room for optimization in the steps outlined above. In the case of CMB-S4, characterizing the ${\sim}500{,}000$ TES bolometers across several hundred wafers will be a significant undertaking \cite{barron22}. Therefore, it is desirable to flag under-performing wafers at room temperature before ever packaging them and installing them in a cryostat. As these larger detector arrays are deployed, this process would reduce time spent testing wafers that have a low probability of being used in an experiment.

Using detector wafers developed for SPT-3G, we attempt to correlate cryogenic performance of TES bolometers with optical images through a random forest machine learning (ML) algorithm. Section \ref{sec:hardware} of these proceedings describes the SPT-3G detector wafers and readout system used. Section \ref{sec:cryo-data} details the capturing of cryogenic data, and Section \ref{sec:imaging} explains the details of our imaging procedure. Section \ref{sec:analysis} covers the details of the analysis procedure and the ML algorithm used to correlate visual and performance data. Lastly, Section \ref{sec:results} covers the results and conclusions of the project.

\section{Detector \& Readout Overview}
\label{sec:hardware}

\subsection{Detectors}
\begin{figure*}
\centering
\includegraphics[width=0.5\columnwidth]{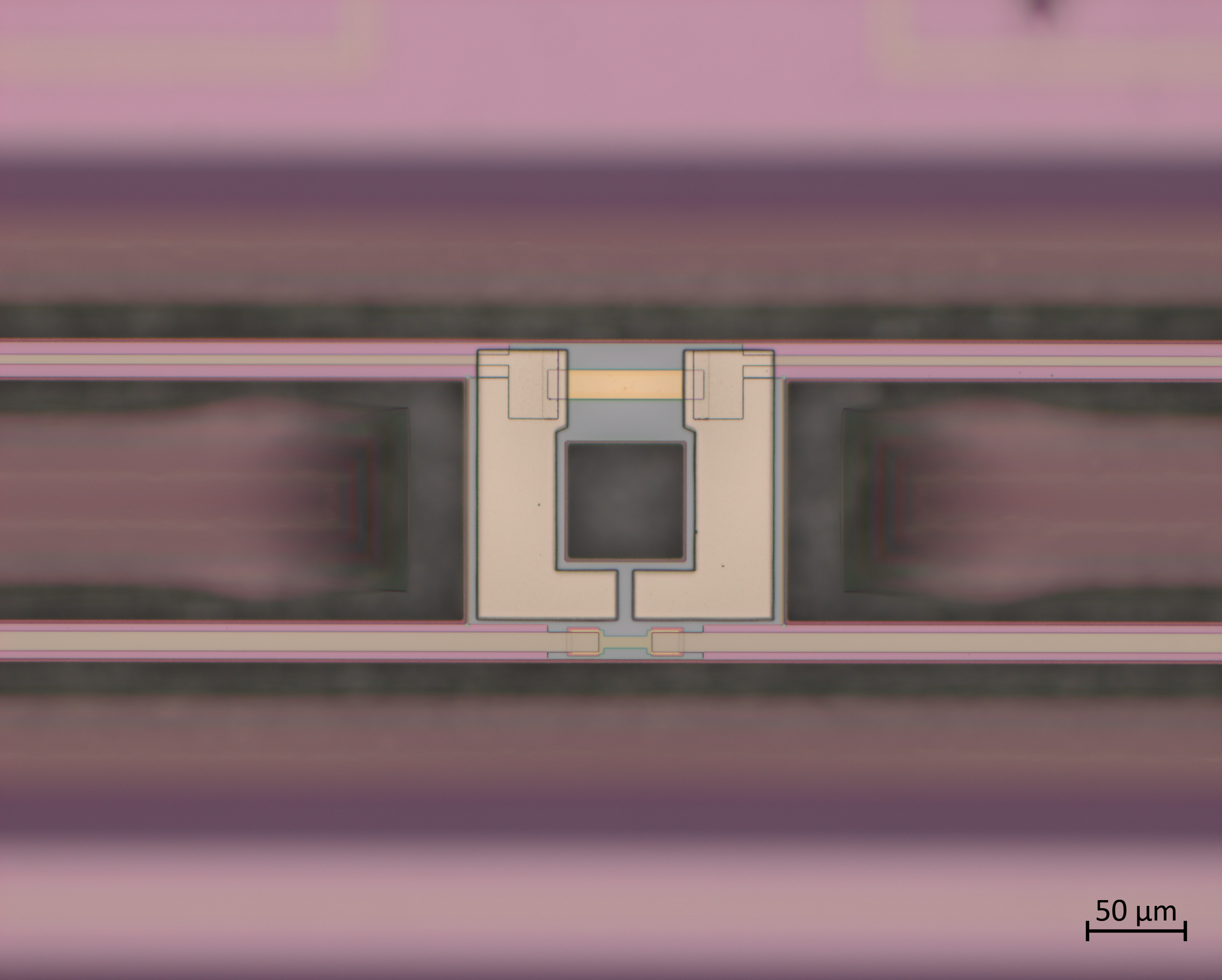}
\caption{A TES bolometer on an SPT-3G detector wafer.}
\label{fig:bolo}
\end{figure*}

For this initial exploration of the ML technique, we use three engineering-grade detector wafers from SPT-3G: W148, W162, and W187. SPT-3G is the third (and current) camera installed on the SPT, and it observes the CMB with $10$ wafers of $1614$ TES detectors each \cite{carter18}. See Ref. \citenum{sobrin22} for further details on the design and operation of SPT-3G. The bolometer architecture was fabricated on a silicon wafer via a multi-step lithographic process \cite{posada15, posada18}; Fig. \ref{fig:bolo} shows an example bolometer. Target TES properties for the SPT-3G fabrication process were:
\begin{itemize}
\item $80 \mu$m long by $15$ or $20 \mu$m wide (the width was changed during the fabrication process, and so different wafers have different nominal TES widths)
\item A normal resistance $R_n = 2.0$ Ohms
\item A critical temperature $T_\textrm{c} = 450$ mK
\item A saturation power $P_\textrm{sat}$ of $10.2$, $15.4$, and $20.0$ pW for detectors in the $95$, $150$, and $220$ GHz observing band, respectively
\end{itemize}

\subsection{Readout}
The SPT-3G hardware utilizes a frequency-domain multiplexing readout system, detailed in Ref. \citenum{bender19} but summarized here. Each detector is placed in series with an $LC$ resonator tuned to a frequency between $1$ and $6$ MHz; up to $68$ of these channels are placed in parallel and read out simultaneously. The signal is amplified in a superconducting quantum interference device (SQUID) and nulled via a feedback loop to keep the SQUID from saturating. The measured outputs are demodulated using the known $LC$ resonant frequencies.

\section{Cryogenic Performance Measurements}
\label{sec:cryo-data}

Data on the cryogenic performance of the detector wafers were captured using a cryogenic chamber with a Simon Chase\footnote{\url{
https://www.chasecryogenics.com/}} He-10 sorption cooler refrigerator backed by a Cryomech\footnote{\url{https://www.cryomech.com/}} PT415 cryocooler. The fridge is capable of reaching temperatures as low as $230$ mK and the cold stage can house one SPT-3G detector wafer at a time. The cryostat is configured for dark measurements; the radiative loading on the detectors is limited by a cover that is held at the base operating temperature.

Once the detectors are cooled down, each readout channel must be mapped to an actual detector on the wafer. We perform such a network analysis by sending in a pure sinusoidal tone whose frequency is slowly swept through the entire possible $LC$ range. Each bolometer has an expected resonant frequency defined by the $LC$ that it is in series with. Expected frequencies are mapped to the measured ones, which are then used for the MHz biases in further detector operation.

\subsection{Detector Tuning}
Once the hardware mapping of all the detector and readout connections in the system is completed, we begin operation by tuning the SQUIDs. They are first heated up to ${\sim} 16$ K to remove any parasitic loop current residing within. After letting them cool, we set a current bias across each SQUID and measure the voltage across the SQUID $V$ as we sweep through a range of flux biases $\phi$. For a well-performing SQUID, this $V{-}\phi$ curve should be a sinusoid; we measure such a curve for a range of current biases, choosing to tune the SQUID to the current bias that yields the largest peak-to-peak amplitude in this sinusoid.

After we've tuned the SQUIDs, we attempt to \textit{overbias} the bolometers. The detectors are brought up to $650$ mK (above the detectors' target $T_\textrm{c}$). While the TES bolometers are in the normal regime, the MHz bias waveforms are initialized, the correct phase of the nulling signal is determined for each channel, and feedback is enabled. At the end of this process, the amplitude of the voltage bias is increased to the level required to maintain the TES in the normal regime. Any bolometer for which this procedure succeeds is labeled \textit{overbiased}.

Lastly, we must tune the bolometers; that is, drop them into their superconducting transition and operate them. The temperature of the detectors is dropped below $T_\textrm{c}$ while maintaining the overbiased state. The bias is then slowly lowered until each detector reaches the target fraction of its normal resistance (usually set to $80 \%$).

\subsection{Detector Performance Characterization}
Once we have verified that we can tune a bolometer, we can proceed to characterizing its performance. We measure two main quantities: resistance $R$ as a function of temperature $T$ and thermal conductance to the bath $G$.

To measure $R(T)$, the bolometers are overbiased at $650$ mK with a minimal amplitude. Using the sorption fridge and supplemental heater, the detectors are slowly swept through a range of temperatures down to a minimum between $300$ and $350$ mK (various minimum temperatures were used to optimize some details of fridge control). Each detector's readout timestream, originally in unitless DAQ counts, is converted to resistance using a factor derived from the readout system \cite{montgomery21}. We measure the same while sweeping the temperature back up to $650$ mK, giving us an up-timestream and a down-timestream for each bolometer.

\begin{figure*}
\centering
\includegraphics[width=0.6\columnwidth]{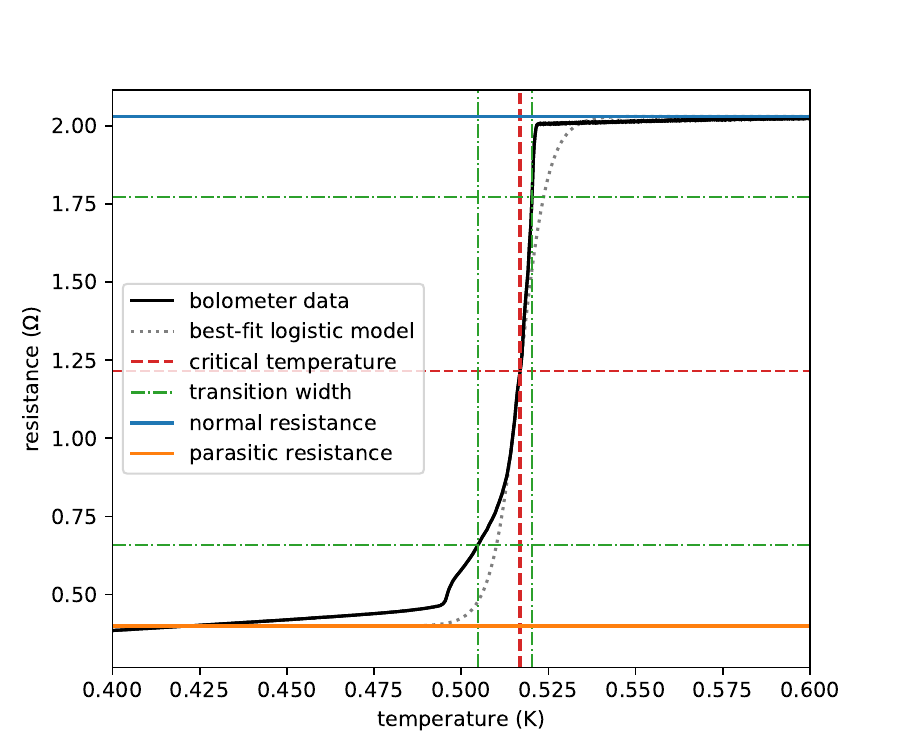}
\caption{Resistance vs. temperature curve for an example bolometer. For each $R(T)$ curve, we determine the best-fit values for the critical temperature $T_\textrm{c}$ (red dotted line), transition width $\Delta T_\textrm{c}$ (dot-dashed green line), normal resistance $R_n$ (blue solid line), and parasitic resistance $R_p$ (orange solid line). The logistic curve used to fit the parameters (Eqn. \ref{eqn:rt_logistic}) is shown as the dotted gray line.}
\label{fig:rt}
\end{figure*}

\begin{figure*}
\centering
\includegraphics[width=0.83\columnwidth]{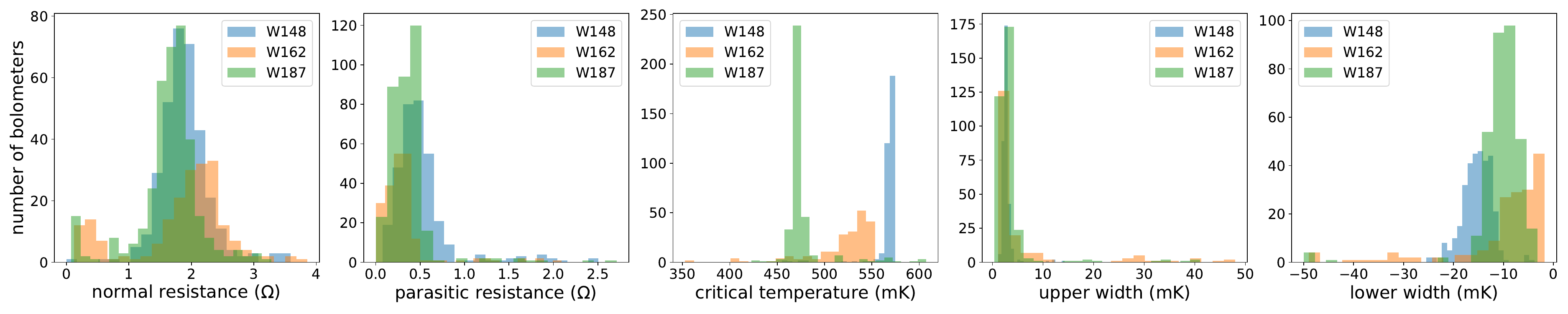}
\caption{Cryogenic features determined from measuring resistance $R$ as a function of temperature $T$.}
\label{fig:output_feats}
\end{figure*}

For each timestream, we determine the best-fit values for $R_n$, $T_\textrm{c}$, the transition width $\Delta T_\textrm{c}$, and the parasitic resistance $R_p$. These parameters are determined by fitting a logistic function of the form

\begin{equation}
\label{eqn:rt_logistic}
R(T) = \frac{R_n}{1 + \exp\left[-d \left( T - T_0 \right) \right]} + R_p
\end{equation}
to the timestream. Here, $T_0$ is the center of the logistic function; it is a possible choice for $T_\textrm{c}$, though for rigor $T_\textrm{c}$ is defined as the temperature where $R(T) = 0.5 \left( R_n - R_p \right) + R_p$. The steepness of the logistic fit is given by $d$; although it is related to $\Delta T_\textrm{c}$, we treat it as a nuisance parameter. Instead, $\Delta T_\textrm{c}$ is defined asymmetrically around $T_\textrm{c}$; the upper bound is the temperature where $R(T) = 0.841\left( R_n - R_p \right) + R_p$ (one standard deviation above center), and the lower bound is the temperature where $R(T) = 0.159\left( R_n - R_p \right) + R_p$ (one standard deviation below center). An example $R(T)$ curve showing the best-fit values is shown in Fig. \ref{fig:rt}. Fig. \ref{fig:output_feats} shows histograms of the measured values for $R_n$, $R_p$, $T_\textrm{c}$, and the upper and lower bounds on $\Delta T_\textrm{c}$, split by wafer.

Data are also captured measuring $G$. This is done by dropping the bolometers into their transition for a range of temperatures between $250$ mK and $550$ mK (for some detectors, the higher temperatures are above their $T_\textrm{c}$; those points are ignored in the analysis). At each temperature, we measure the saturation power $P_\textrm{sat}$. For simplicity, information about $P_\textrm{sat}$ and $G$ is left out of this analysis and saved for future work.

\section{Imaging Procedure}
\label{sec:imaging}

Optical images of the detectors were captured on a Zeiss Axio Imager Vario microscope housed at the Center for Nanoscale Materials at Argonne National Laboratory. The imaging process, including moving to each detector's location, focusing, and capturing the image, was fully automated using the Zeiss ZEN Microscopy Software. Although the highest objective lens on this piece of equipment is $100$x magnification, we elected to use the $20$x objective lens as this allowed easier capture of the entire bolometer island architecture \cite{sobrin22}. Resulting images were $2752$ pixels wide by $2208$ pixels tall, with a scale factor of $0.227$ microns per pixel. Images were captured in full color (see Fig. \ref{fig:bolo}), although only their grayscale information was used for this analysis.

\begin{figure*}
\centering
\includegraphics[width=0.6\columnwidth]{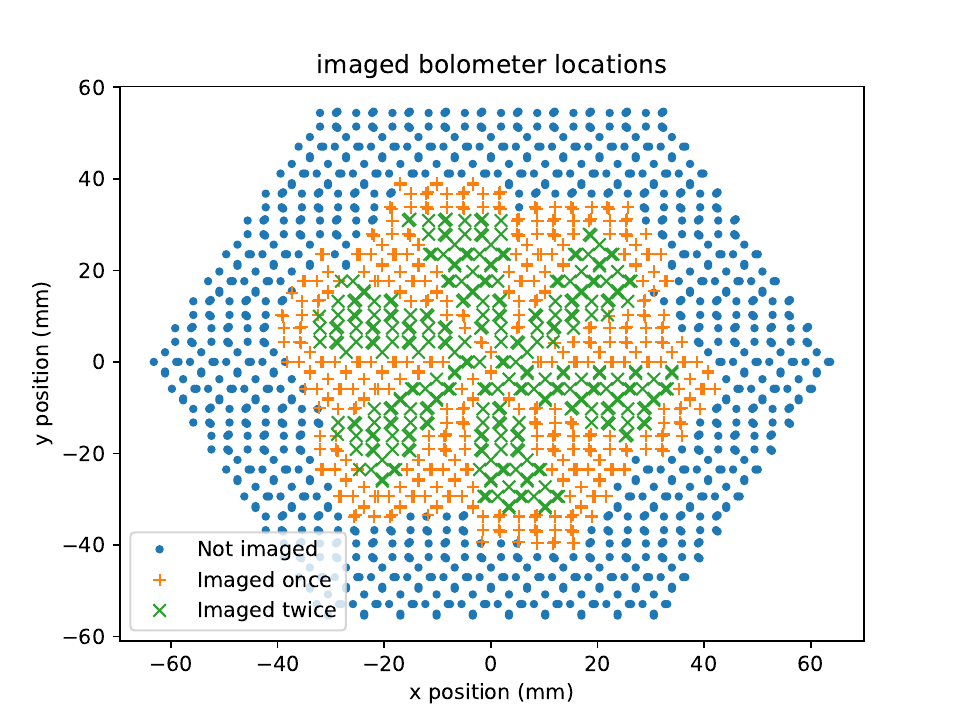}
\caption{Bolometer locations relative to wafer center. Bolometers represented by blue dots were not imaged due to their proximity to the wafer edge. Bolometers represented by orange plus signs were imaged once and those represented by green exes were imaged twice.}
\label{fig:bolo_locs}
\end{figure*}

In the ideal case, images would be captured for every detector on a wafer. However, the SPT-3G detector wafers used here were packaged into modules, ready for cryogenic characterization. Part of the module housing is a holding structure allowing them to be mounted in the refrigerator. It was impossible to remove all of this structure without breaking the wirebonds connecting the wafer to the cables that attach to the readout $LC$ chips. Due to the size of the objective lens, we could not image detectors near the edge of the wafer without colliding with the remaining housing structure; thus, boundaries had to be set on where we could take images. These boundaries were set conservatively to avoid all possibility of a collision. In total we were able to image about $44\%$ of the detectors (that is, ${\sim} 713$ detectors) on each wafer. Due to the limited available imaging region, images were captured in six groups, with the wafer physically rotated underneath the microscope between each group. These groups partially overlapped, causing $42 \%$ of the imaged detectors (${\sim} 300$ detectors per wafer) to be imaged twice. The locations of the singly- and doubly-imaged detectors are shown in Fig. \ref{fig:bolo_locs}.

Additional challenges are inherent to the imaging process. The bolometer island is suspended over a trough as a result of the fabrication process. Occasionally, the microscope autofocus procedure will focus on this trough rather than the bolometer itself. This reduces our usable data volume, but only by $\lesssim 10$ bolometers per wafer. Another consequence of this geometry is that some bolometers ($\lesssim 5\%$ on a single wafer) are not quite flat relative to the rest of the wafer. This is due to differing tension in the legs holding up the bolometer island on either side. For these detectors, the level of focus is a gradient across the island. In most cases, this adds some variance to the features we calculate from the images (Section \ref{sec:analysis}), though in extreme cases it can lead to a failure to calculate any features at all.

Because the details of the bolometer island legs affect the thermal conductance $G$ of the bolometer, we expect that the severity of this gradient will correlate with $G$. As previously stated, this consideration is left for future work.

\section{Analysis}
\label{sec:analysis}

In an ideal scenario, a machine learning algorithm would train directly on the images described in Section \ref{sec:imaging}. However, this proved unfeasible with the amount of data available. While there are $\mathcal{O}\left( 1{,}000{,}000 \right)$ pixels per image, images were only captured for $\mathcal{O}\left( 1{,}000 \right)$ bolometers across the three wafers. Training directly on the images would yield an algorithm prone to overfitting, and one that would likely do a poor job estimating the properties of bolometers that were not used in the training set. For this reason, it was necessary to determine a set of features that could be calculated from the images and use those as inputs to the ML algorithm.

\subsection{Image Feature Determination}

\begin{figure*}
\centering
\includegraphics[width=0.668\columnwidth]{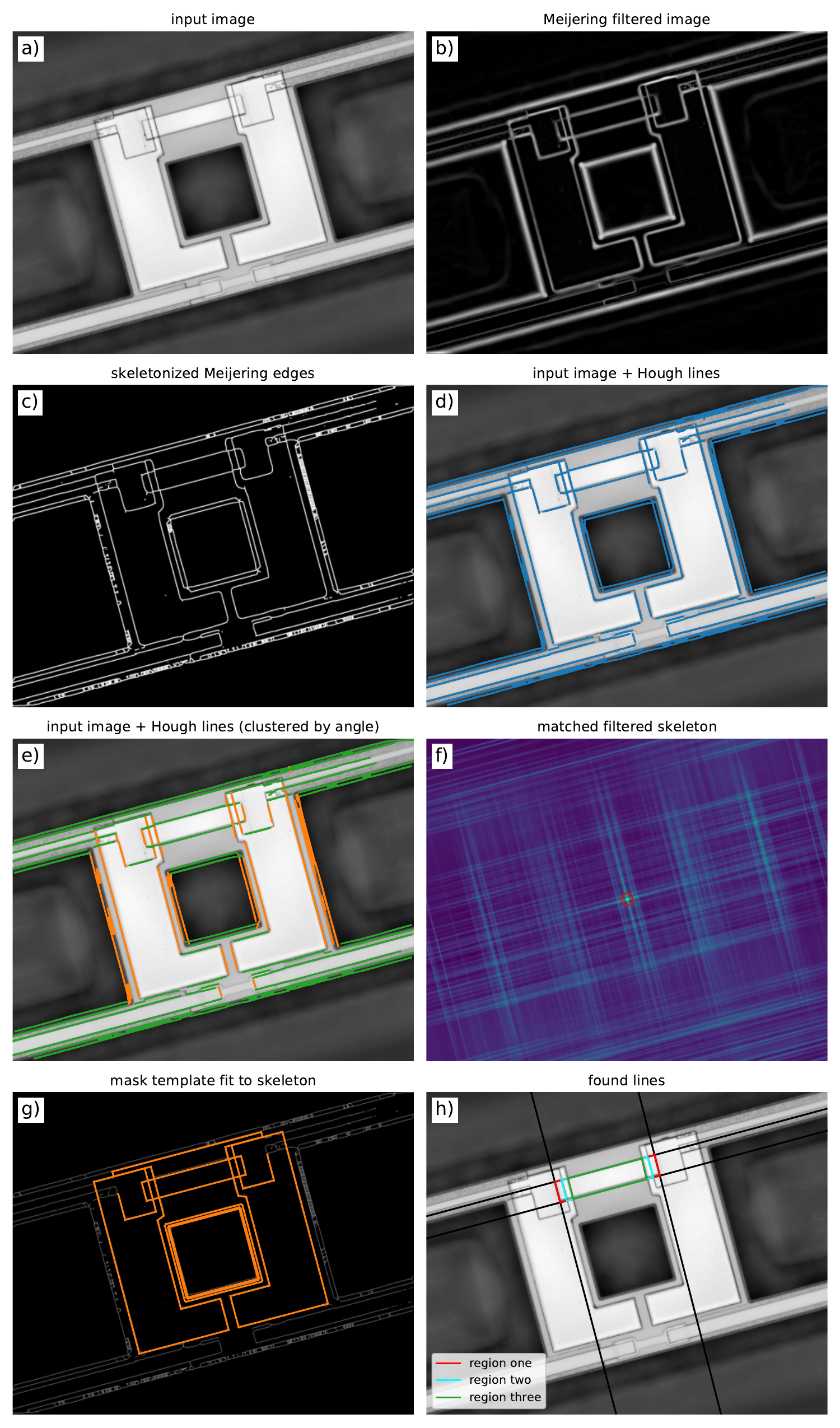}
\caption{Step-by-step process of determining the location of the TES in an image. a) The image is trimmed around the center and converted to grayscale. b) The image is filtered with a Meijering edge detection algorithm. c) The Meijering edges are skeletonized to become a single pixel wide. d) Probabilistic Hough lines are calculated from the skeleton. e) The Hough lines are divided by angle into two groups to determine the overall angle of the TES. f) A template of the most prominent lines on the bolometer island is matched filtered with the skeleton to find the exact location and orientation. g) Best-fit template overlaid on the skeleton. h) Final lines found for TES edges, as well as overlap with other parts of the bolometer island.}
\label{fig:location}
\end{figure*}

All of the derived features require determining where the TES lies in the image. Due to varying orientation angle, drift of the wafer relative to the stage during the imaging process, and slight fabrication variance, this is not as trivial as simply picking out the same pixels in every image. The process for determining the TES location is shown step-by-step in Fig. \ref{fig:location}. We first cut out a patch from the center of the image of size $1376 \times 1104$ pixels (half the size of the original image in each dimension) and convert the image to grayscale (Fig. \ref{fig:location}a). We then apply a Meijering edge detection filter to the image (Fig. \ref{fig:location}b) \cite{meijering04}. The resulting array is binarized, with all pixels whose values are less than $20\%$ of the maximum set to zero, and all other pixels set to one.

This binary mask is \textit{skeletonized} (Fig. \ref{fig:location}c), a process by which all of the detected edges are converted to be a single pixel wide (the width of the edges has been artificially increased in Fig. \ref{fig:location}c for visibility). The skeleton can then be directly compared to a template (generated from the fabrication layout) of the most prominent lines on the bolometer island. The location and orientation angle of the template that give the highest correlation with the skeleton are used to determine the TES location in the image from its location in the template. First, we determine the angle by taking a probabilistic Hough transform \cite{galamhos99} of the skeleton (Fig. \ref{fig:location}d). Since the resulting line segments are mostly orthogonal or parallel to each other, we can divide them into two groups by their angle (Fig. \ref{fig:location}e). We expect more lines in the horizontal direction (relative to the TES) than the vertical, so this gives us the orientation angle of the template up to a $180$-degree modulus. For the two remaining possible orientation angles, we perform a matched filter between the template and the skeleton (Fig. \ref{fig:location}f). The location of the maximum-valued pixel gives us both the offset from center and the orientation angle of the template.

To identify the line segments corresponding to the TES edges, we must have some foreknowledge of where we expect the TES to be. As a consequence of the matched filter, we already know this; since we know which lines in the template correspond to the TES edges, we then have an expected location within the image for the TES. We construct a $15$-pixel buffer region around each expected line location and select the longest Hough line in said region to represent the true edge in the image. In addition to the TES edges, we identify Hough lines corresponding to the lines where the bling and lead intersect with the TES. These found lines define three regions, which are shown in Fig. \ref{fig:location}h. Because the probabilistic Hough transform is random by nature, it returns a different set of line segments every time it is run on an identical skeleton. We run this process on each image $100$ times and use the average vertex location to calculate our features.

$14$ features are calculated per image:
\begin{itemize}
\item We calculate the area and perimeter of the TES (\textit{region one}, red lines in Fig. \ref{fig:location}h), as well as the areas and perimeters of the smaller regions not covered by the lead (\textit{region two}, aqua lines in Fig. \ref{fig:location}h) and the bling (\textit{region three}, green lines in Fig. \ref{fig:location}h).
\item We estimate the roughness of the detector surface by calculating the mean and standard deviation of the pixel values within region three.
\item Additionally, we quantify the roughness of the TES edges. This can be affected by detritus accumulating along the TES edges or by slight fabrication non-idealities. To estimate the roughness, we identify the pixels at which the Meijering-filtered image takes on local maxima in the vicinity of the TES. We measure the number of these pixels as well as the mean and standard deviation of their values (these quantities were shown to roughly correlate with the roughness of the TES edges in visual inspection).
\item Occasionally, pieces of the bolometer architecture (such as the bling that serves to dissipate heat) are missing or obscured in the image. In these cases, we expect the matched filter between the template and the skeletonized image to return a smaller maximum value. To provide information on bolometers where this happens, we take this maximum value as a feature.
\item Because some of the metrics described above are dependent on pixelization effects that differ at different angles, we also include the orientation angle of the TES as a feature.
\item Lastly, occasional failures of the location-finding algorithm (usually due to either poor focus or missing pieces of the bolometer architecture) will lead to missing values for some or all features on a given detector. These values are imputed using the MissForest algorithm \cite{stekhoven12}, but in order to include some information about these failures we include a binary flag for whether there were any missing values before the imputation.
\end{itemize}

\begin{figure*}
\centering
\includegraphics[width=1.0\columnwidth]{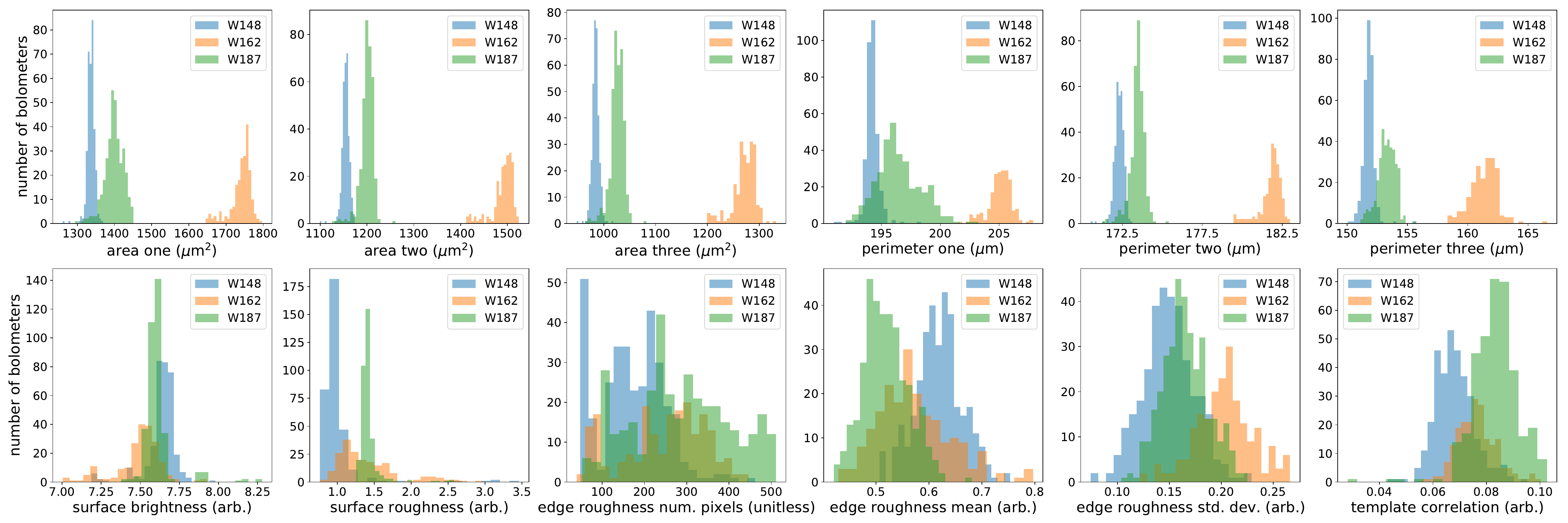}
\caption{Image features (except for the TES angle and the flag for missing features) split by wafer. The label \textit{one} (\textit{two}) [\textit{three}] for the area and perimeter features represents the full TES (region of the TES not overlapped by the lead) [region of the TES not overlapped by anything else]; see Fig. \ref{fig:location}h for visual clarification. \textit{Surface brightness/roughness} are the mean/standard deviation of the values of the pixels in region three. \textit{Edge roughness} denotes the number of local maxima pixels, as well as their mean and standard deviation, for region one of the Meijering-filtered image. \textit{Template correlation} denotes the maximum value of the matched-filtered skeleton with the bolometer island template.}
\label{fig:input_feats}
\end{figure*}

As discussed in Section \ref{sec:imaging}, a subset of bolometers were imaged twice at different angles. For each subgroup of data on which we train an ML algorithm, two algorithms are actually trained: one in which the features from the doubly-imaged bolometers are kept, and one in which they are averaged. In the case where they are averaged, \textit{all} features are averaged, even the angle and the binary flag for whether there were missing values before the data imputation. Fig. \ref{fig:input_feats} histograms the value of all input features (except the angle and the binary flag). In these histograms, features from doubly-imaged bolometers are averaged and values imputed using MissForest are not included.

\subsection{Machine Learning Algorithm}
There are two main predictors of a wafer's experimental viability: its detector yield and the performance of those detectors. We wish to predict both of these. For the first, we train a random forest classification algorithm; for the second, a random forest regression algorithm \cite{breiman01}. In both cases, we train on multiple subgroups of data:
\begin{itemize}
\item Individual wafer performance ($80/20$ train/test split)
\item Train on two wafers, test the third
\item Train on all wafers, with a random $20\%$ of bolometers chosen for testing
\end{itemize}

Every time an algorithm is trained, algorithm hyperparameters are decided via an exhaustive grid search. Four hyperparameters are varied: the number of trees in the random forest $N_\textrm{trees} \in \left[ 10, 50,100, 500, 1000 \right]$, the maximum depth of each tree $d \in \left[ 1,2,3,4,5,6 \right]$\footnote{Initially, unlimited tree depth was an option. However, examination of algorithm learning curves determined that this led to overfitting and so the option was removed.}, the minimum number of bolometers required to split an internal node on a tree $s \in \left[ 2,3,4 \right]$, and the fraction of bolometers (with replacement) used to train each tree $f_\textrm{bolos} \in \left[ 0.1,0.25,0.5,1.0 \right]$. For each combination of hyperparameters, the algorithm performance is validated via five-fold cross validation \cite{stone74}. The training set is randomly split into five subgroups; each subgroup is in turn left out of training and used to determine a performance score, with the overall score for that combination of hyperparameters being the mean of the scores for all five subgroups. The final sets of hyperparameters used for each predictor are listed in the tables in Appendix \hyperref[sec:appendix]{A}.

To classify detector yield, the bolometers are grouped into one of three categories: fully tuneable, able to overbias but not tune, and able to do neither. Because the percentage of bolometers in each class is not equal, we specifically perform a stratified cross validation here in which the ratio of bolometer classes is retained when splitting the training set into subgroups during cross validation. A simple classification accuracy was used to score performance.

Six output features $y$ were used to characterize detector performance: $R_n$, $R_p$, $T_\textrm{c}$, and the one-standard-deviation-up and -down temperatures described in Section \ref{sec:cryo-data}. As with the input features calculated from the images, some of these features are occasionally missing for a given bolometer. Missing values are imputed using the MissForest algorithm, and an additional feature flagging on whether there were any missing features before the imputation is included. The training data is scored with the $R^2$ coefficient, $R^2 \equiv 1 - \frac{u}{v}$, where $u = \sum_i \left( y_\textrm{true, i} - y_\textrm{predicted, i} \right)^2$ and $v = \sum_i \left( y_\textrm{true, i} - \bar{y}_\textrm{true, i} \right)^2$, and where the bar signifies the mean over the test set. Each output feature is scored separately, and the overall score is given by the mean of the scores in all six features.

Interpreting the training scores of the various algorithms requires the context of how they would perform if there were no correlation between the input features and the output classes/features. In order to quantify this, we shuffle the output classes/features in both the training and test sets and re-train/re-test the random forest on the shuffled data. The same set of input features is used, and the training/test sets are shuffled together. This is done $100$ times and the mean and standard deviation of the resultant scores are used to estimate the no-skill performance of the algorithm.

\section{Results \& Discussion}
\label{sec:results}

\begin{table*}
\centering
\caption{Random forest classification performance results. The \textit{True} scores are from the actual data while the \textit{Shuffled} scores are the mean plus/minus the standard deviation of $100$ training/test iterations where the output classes were randomly shuffled. All predictors use an 80/20 training/test split unless stated otherwise.}
\def\arraystretch{1.5}
\setlength{\tabcolsep}{10pt}
\begin{tabular}{ c|c|c|c }
\hline\hline
 \multicolumn{4}{c}{Classification Accuracy} \\
 \hline
 \multicolumn{2}{c|}{\multirow{2}{*}{ML Predictor}} & \multicolumn{2}{c}{Features for bolometers with two images} \\
 \cline{3-4}
 \multicolumn{2}{c|}{} & Kept & Averaged \\
 \hline\hline
 \multirow{2}{*}{W148 only} & True & 0.6281 & 0.6627 \\
 \cline{2-4}
 & Shuffled & 0.6358 $\pm$ 0.0356 & 0.6199 $\pm$ 0.0472 \\
 \hline
 \multirow{2}{*}{W162 only} & True & 0.7013 & 0.6852 \\
 \cline{2-4}
 & Shuffled & 0.6978 $\pm$ 0.0466 & 0.6669 $\pm$ 0.0568 \\
 \hline
 \multirow{2}{*}{W187 only} & True & 0.5430 & 0.5673 \\
 \cline{2-4}
 & Shuffled & 0.4797 $\pm$ 0.0375 & 0.4991 $\pm$ 0.0471 \\
 \hline
 \multirow{2}{*}{Train W148+W162, Test W187} & True & 0.5258 & 0.5367 \\
 \cline{2-4}
 & Shuffled & 0.6034 $\pm$ 0.0142 & 0.5990 $\pm$ 0.0155 \\
 \hline
 \multirow{2}{*}{Train W148+W187, Test W162} & True & 0.5459 & 0.2772 \\
 \cline{2-4}
 & Shuffled & 0.5292 $\pm$ 0.1201 & 0.5624 $\pm$ 0.0828 \\
 \hline
 \multirow{2}{*}{Train W162+W187, Test W148} & True & 0.3261 & 0.2530 \\
 \cline{2-4}
 & Shuffled & 0.5934 $\pm$ 0.0363 & 0.5599 $\pm$ 0.0689 \\
 \hline
 \multirow{2}{*}{All wafers, random train/test} & True & 0.6178 & 0.6083 \\
 \cline{2-4}
 & Shuffled & 0.6013 $\pm$ 0.0226 & 0.5996 $\pm$ 0.0289 \\
 \hline\hline
\end{tabular}
\label{tab:results_classification}
\end{table*}

\begin{table*}
\centering
\caption{Random forest regression performance results. The \textit{True} scores are from the actual data while the \textit{Shuffled} scores are the mean plus/minus the standard deviation of $100$ training/test iterations where the output classes were randomly shuffled. All predictors use an 80/20 training/test split unless stated otherwise.}
\def\arraystretch{1.5}
\setlength{\tabcolsep}{10pt}
\begin{tabular}{ c|c|c|c }
\hline\hline
 \multicolumn{4}{c}{$R^2$ score} \\
 \hline
 \multicolumn{2}{c|}{\multirow{2}{*}{ML Predictor}} & \multicolumn{2}{c}{Features for bolometers with two images} \\
 \cline{3-4}
 \multicolumn{2}{c|}{} & Kept & Averaged \\
 \hline\hline
 \multirow{2}{*}{W148 only} & True & -0.0216 & -0.0292 \\
 \cline{2-4}
 & Shuffled & -0.0161 $\pm$ 0.0168 & -0.0755 $\pm$ 0.1198 \\
 \hline
 \multirow{2}{*}{W162 only} & True & -0.0040 & -0.0182 \\
 \cline{2-4}
 & Shuffled & -0.0328 $\pm$ 0.0317 & -0.0521 $\pm$ 0.0534 \\
 \hline
 \multirow{2}{*}{W187 only} & True & -0.0335 & -0.0183 \\
 \cline{2-4}
 & Shuffled & -0.0249 $\pm$ 0.0240 & -0.0400 $\pm$ 0.0502 \\
 \hline
 \multirow{2}{*}{Train W148+W162, Test W187} & True & -1.5543 & -1.6732 \\
 \cline{2-4}
 & Shuffled & -0.0195 $\pm$ 0.0205 & -0.0196 $\pm$ 0.0140 \\
 \hline
 \multirow{2}{*}{Train W148+W187, Test W162} & True & -0.1947 & -0.2316 \\
 \cline{2-4}
 & Shuffled & -0.1141 $\pm$ 0.2090 & -0.1761 $\pm$ 0.2760 \\
 \hline
 \multirow{2}{*}{Train W162+W187, Test W148} & True & -10.3544 & -8.8573 \\
 \cline{2-4}
 & Shuffled & -0.0548 $\pm$ 0.0514 & -0.0282 $\pm$ 0.0287 \\
 \hline
 \multirow{2}{*}{All wafers, random train/test} & True & 0.1207 & 0.1046 \\
 \cline{2-4}
 & Shuffled & -0.0175 $\pm$ 0.0099 & -0.0180 $\pm$ 0.0125 \\
 \hline\hline
\end{tabular}
\label{tab:results_regression}
\end{table*}

\begin{figure*}
\centering
\includegraphics[width=0.75\columnwidth]{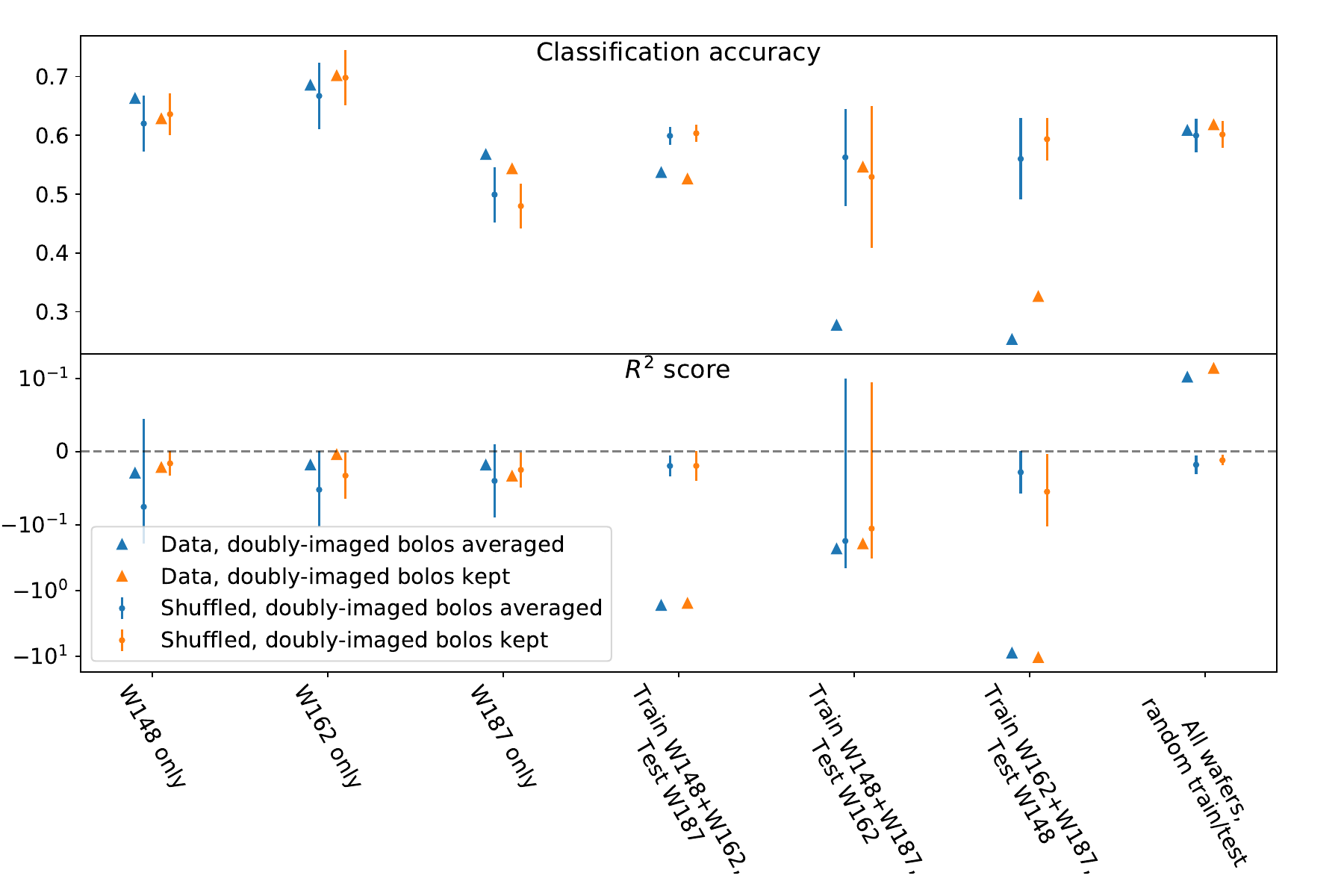}
\caption{Visualization of the results displayed in Tables \ref{tab:results_classification} and \ref{tab:results_regression}.}
\label{fig:results}
\end{figure*}

The results for the random forest classification are shown in Table \ref{tab:results_classification}, and the results for the random forest regression in Table \ref{tab:results_regression}. The rows labeled \textit{True} give the scores from the real training/test data, while the rows labeled \textit{Shuffled} give the mean and the standard deviation of the scores of the shuffled data as described in Section \ref{sec:analysis}. These results are visualized in Fig. \ref{fig:results}.

Generally, the results are fairly insensitive to the specific treatment of doubly-imaged bolometers. The main exception is in the classification accuracies for the \textit{Train W148+W187, Test W162} predictor. However, given the size of the \textit{Shuffled} error bars and the similarity between \textit{Kept}/\textit{Averaged} cases in other predictors, it is likely this is due to statistical scatter rather than a real effect.

In most cases, a predictor's \textit{True} score is not meaningfully different from its mean \textit{Shuffled} score, implying that the random forest lacks predictive power. However, the \textit{True} scores are significantly lower than the mean \textit{Shuffled} scores for many of the predictors where the algorithm is trained on data from two wafers and tested on data from a third. This discrepancy is moderate for the classification problem, but especially significant for the regression problem. This behavior arises because many of the input features are tri-modal (see Fig. \ref{fig:input_feats}); each wafer exhibits its own behavior (even among those with supposedly identical fabrication specifications), and features vary much more wafer-to-wafer than within individual wafers. Random forests do not perform well on data with input features outside of the training set, hence the poor predictions.

The most statistically-significant instance of a predictor performing better than random guessing is the \textit{All wafers, random test/train} regression predictor. This is also the only regression predictor to have a positive score; that is, to do better than it would had it simply guessed the test set mean for every feature. It is tempting to ascribe this increase in performance to the fact that more data are used to train the algorithm. However, due to the tri-modality described above, the random forest is essentially predicting which wafer a bolometer is on (with high precision) and assigning it the mean output features from that wafer. While having this predictive power indicates that the random forest algorithm is in some way performing as it should, it is less useful for real-world applications, where the goal is to accurately predict cryogenic features of detectors on wafers that were not used in the training set. This is perhaps not surprising; given that feature variation is much larger between wafers than within them, our effective sample size is much reduced. Many more wafers would be needed to begin predicting full-wafer features.

Future work on this topic can expand on the current study in a number of ways:
\begin{itemize}
\item As with any ML problem, the collection of more data is likely to yield better results with stronger predictive power.
\item This study has focused on only a limited amount of the visual information available in the bolometer images. Calculating more input features (or even finding a way to train on the images themselves while still avoiding overfitting) could prove informative.
\item This study used an ``out-of-the-box'' measure of classification accuracy. Changing the decision threshold by optimizing precision/recall (or, equivalently, the $F_1$ score) in a one-vs-all or one-vs-one classification scheme could yield better performance.
\item Bolometer performance depends on more than the limited set of output features predicted in the regression algorithms. Future work should incorporate information on the thermal conductance $G$, the saturation power $P_\textrm{sat}$, and perhaps even optical bolometer properties.
\end{itemize}

Although it is unclear how much data must be included in the training set before the random forests that were trained gain predictive power on new detector wafers, the algorithms were extremely successful at predicting which wafer a bolometer belonged to. Furthermore, the image analysis proved adept at revealing previously unknown defects. Even without predictive power from a machine learning algorithm, the image analysis holds potential for identifying potentially flawed detectors prior to cryogenic characterization. Finally, we note that the image analysis is easily generalized to other detector geometries for use in future experiments.

\section*{APPENDIX A}
\label{sec:appendix}
Here we include the optimal set of hyperparameters that were determined for each of the random forests that were trained. Table \ref{tab:hyperparams_Classification} lists those determined for the classification algorithms, and Table \ref{tab:hyperparams_Regression} the regression algorithms.

\begin{table*}[bp]
\centering
\caption{Optimal random forest hyperparameters for classification. $N_\textrm{trees}$ is the number of trees in the random forest; choices were $\left[10, 50, 100, 500, 1000 \right]$. $d$ is the maximum allowed depth of each tree; choices were $\left[1, 2, 3, 4, 5, 6 \right]$. $s$ is the minimum number of bolometers necessary to split a node on the tree; choices were $\left[ 2, 3, 4 \right]$. $f_\textrm{bolos}$ is the fraction of the total number of bolometers used to train each tree, chosen with replacement from the full set; choices were $\left[ 0.1, 0.25, 0.5, 1.0 \right]$.}
\def\arraystretch{1.5}
\setlength{\tabcolsep}{10pt}
\begin{tabular}{ c|c|c|c|c|c }
\hline\hline
 \multicolumn{6}{c}{Classification Hyperparameters} \\
 \hline
 \multirow{2}{*}{ML Predictor} & \multirow{2}{0.22\linewidth}{\centering Features for bolometers with two images} & \multicolumn{4}{c}{Hyperparameters} \\
 \cline{3-6}
 & & $N_\textrm{trees}$ & $d$ & $s$ & $f_\textrm{bolos}$ \\
 \hline\hline
 \multirow{2}{0.18\linewidth}{\centering W148 only} & Kept & 10 & 4 & 2 & 0.25 \\
 \cline{2-6}
 & Averaged & 50 & 4 & 3 & 0.25 \\
 \hline
 \multirow{2}{0.18\linewidth}{\centering W162 only} & Kept & 10 & 2 & 3 & 1.0 \\
 \cline{2-6}
 & Averaged & 10 & 2 & 2 & 1.0 \\
 \hline
 \multirow{2}{0.18\linewidth}{\centering W187 only} & Kept & 10 & 6 & 2 & 0.25 \\
 \cline{2-6}
 & Averaged & 10 & 5 & 2 & 0.5 \\
 \hline
 \multirow{2}{0.18\linewidth}{\centering Train W148+W162, Test W187} & Kept & 100 & 6 & 4 & 0.25 \\
 \cline{2-6}
 & Averaged & 50 & 6 & 3 & 0.1 \\
 \hline
 \multirow{2}{0.18\linewidth}{\centering Train W148+W187, Test W162} & Kept & 50 & 6 & 2 & 1.0 \\
 \cline{2-6}
 & Averaged & 500 & 6 & 2 & 0.25 \\
 \hline
 \multirow{2}{0.18\linewidth}{\centering Train W162+W187, Test W148} & Kept & 500 & 6 & 4 & 0.5 \\
 \cline{2-6}
 & Averaged & 50 & 6 & 4 & 0.25 \\
 \hline
 \multirow{2}{0.18\linewidth}{\centering All wafers, random train/test} & Kept & 10 & 5 & 4 & 1.0 \\
 \cline{2-6}
 & Averaged & 50 & 6 & 4 & 1.0 \\
 \hline\hline
\end{tabular}
\label{tab:hyperparams_Classification}
\end{table*}

\begin{table*}
\centering
\caption{Optimal random forest hyperparameters for classification. $N_\textrm{trees}$ is the number of trees in the random forest; choices were $\left[10, 50, 100, 500, 1000 \right]$. $d$ is the maximum allowed depth of each tree; choices were $\left[1, 2, 3, 4, 5, 6 \right]$. $s$ is the minimum number of bolometers necessary to split a node on the tree; choices were $\left[ 2, 3, 4 \right]$. $f_\textrm{bolos}$ is the fraction of the total number of bolometers used to train each tree, chosen with replacement from the full set; choices were $\left[ 0.1, 0.25, 0.5, 1.0 \right]$.}
\def\arraystretch{1.5}
\setlength{\tabcolsep}{10pt}
\begin{tabular}{ c|c|c|c|c|c }
\hline\hline
 \multicolumn{6}{c}{Regression Hyperparameters} \\
 \hline
 \multirow{2}{*}{ML Predictor} & \multirow{2}{0.22\linewidth}{\centering Features for bolometers with two images} & \multicolumn{4}{c}{Hyperparameters} \\
 \cline{3-6}
 & & $N_\textrm{trees}$ & $d$ & $s$ & $f_\textrm{bolos}$ \\
 \hline\hline
 \multirow{2}{0.18\linewidth}{\centering W148 only} & Kept & 50 & 1 & 3 & 1.0 \\
 \cline{2-6}
 & Averaged & 10 & 1 & 4 & 0.5 \\
 \hline
 \multirow{2}{0.18\linewidth}{\centering W162 only} & Kept & 50 & 1 & 2 & 0.25 \\
 \cline{2-6}
 & Averaged & 100 & 1 & 3 & 0.1 \\
 \hline
 \multirow{2}{0.18\linewidth}{\centering W187 only} & Kept & 10 & 3 & 3 & 1.0 \\
 \cline{2-6}
 & Averaged & 10 & 1 & 4 & 0.5 \\
 \hline
 \multirow{2}{0.18\linewidth}{\centering Train W148+W162, Test W187} & Kept & 100 & 5 & 2 & 1.0 \\
 \cline{2-6}
 & Averaged & 100 & 5 & 2 & 1.0 \\
 \hline
 \multirow{2}{0.18\linewidth}{\centering Train W148+W187, Test W162} & Kept & 50 & 2 & 3 & 1.0 \\
 \cline{2-6}
 & Averaged & 100 & 3 & 4 & 1.0 \\
 \hline
 \multirow{2}{0.18\linewidth}{\centering Train W162+W187, Test W148} & Kept & 500 & 5 & 2 & 1.0 \\
 \cline{2-6}
 & Averaged & 1000 & 3 & 3 & 1.0 \\
 \hline
 \multirow{2}{0.18\linewidth}{\centering All wafers, random train/test} & Kept & 1000 & 6 & 3 & 1.0 \\
 \cline{2-6}
 & Averaged & 500 & 6 & 2 & 1.0 \\
 \hline\hline
\end{tabular}
\label{tab:hyperparams_Regression}
\end{table*}

\section*{ACKNOWLEDGMENTS}
\markboth{ACKNOWLEDGMENTS}{ACKNOWLEDGMENTS}
The authors would like to thank Claudio Kopper, Matthew Becker, Nesar Ramachandra, and Markus Rau for their helpful conversations, as well as the SPT-3G collaboration for the use of their detector wafers for this study. The South Pole Telescope program is supported by the National Science Foundation (NSF) through grants PLR-1248097, OPP-1852617. Work at Argonne, including use of the Center for Nanoscale Materials, an Office of Science user facility, was supported by the U.S. Department of Energy, Office of Science, Office of Basic Energy Sciences and Office of High Energy Physics, under Contract No. DE-AC02- 06CH11357. We gratefully acknowledge the computing resources provided on Crossover, a high-performance computing cluster operated by the Laboratory Computing Resource Center at Argonne National Laboratory. KF acknowledges support from the U.S. Department of Energy's Office of Science Graduate Student Research (SCGSR) Program. This work makes use of the \texttt{numpy} \cite{harris20}, \texttt{matplotlib} \cite{hunter07}, \texttt{scipy} \cite{virtanen20}, \texttt{scikit-image} \cite{vanderwalt14}, and \texttt{scikit-learn} \cite{pedregosa11} Python packages.

\bibliography{refs}
\bibliographystyle{spiebib} 

\end{document}